\documentclass[a4paper]{jpconf}
\usepackage{graphicx}
\usepackage{bm}
\usepackage{hyperref}
\hypersetup{colorlinks,linkcolor=blue,filecolor=green,urlcolor=blue,citecolor=blue}

\usepackage{dcolumn}
\usepackage{mathrsfs}
\usepackage{amsmath}
\usepackage{graphicx}
\usepackage{bm}
\usepackage{epstopdf}
\usepackage{float}
\usepackage{hyperref}
\usepackage{color}
\usepackage{amssymb}
\usepackage{eucal}
\usepackage{mathrsfs}
\usepackage{amsthm}
\usepackage{braket}

\def\br{{\bf r}}

\def\bk{{\bf k}}

\def\bm{{\boldsymbol \mu}}
\def\bE{{\bf E}}
\def\w0{\omega_0}
\def\k0{k_0}
\def\wk{\omega_k}

\def\akj{a_{\bk j}}
\def\akjd{a_{\bk j}^\dagger}

\begin{document}

\title{Dynamical atom-wall Casimir-Polder effect after a sudden change of the atomic position}

\author{Antonio Noto$^1$, Roberto Passante$^{1,2}$, Lucia Rizzuto$^{1,2}$, Salvatore Spagnolo$^1$}

\address{$^1$Dipartimento di Fisica e Chimica - Emilio Segrè, Università degli Studi di Palermo, Via Archirafi 36, I-90123 Palermo, Italia}
\address{$^2$ INFN, Laboratori Nazionali del Sud, I-95123 Catania, Italia}
\ead{lucia.rizzuto@unipa.it}

\begin{abstract}
We investigate the dynamical Casimir-Polder force between an atom and a conducting wall during the time evolution of the system from a partially dressed state. This state is obtained by a sudden change of the atomic position with respect to the plate. To evaluate the time-dependent  atom-plate Casimir-Polder force we solve the Heisenberg equations for the field and atomic operators by an iterative technique. We find that the dynamical atom-plate Casimir-Polder interaction exhibits oscillation in time, and can be attractive or repulsive depending on time and the atom-wall distance. We also investigate the time dependence of global observables, such as the field and atomic Hamiltonians, and discuss some interesting features of the dynamical process bringing the interaction energy to the equilibrium configuration. 
\end{abstract}

\section{Introduction}
Vacuum field fluctuations of the quantized electromagnetic field are a consequence of the Heisenberg uncertainty relations and are at the origin of many physical phenomena, for example the Lamb shift \cite{Lamb-Retherford47, Maclay20}, the spontaneous emission by an excited atom \cite{Milonni19}, the resonant and dispersion Casimir-Polder  interactions between neutral atoms \cite{Casimir48, Casimir-Polder48, Compagno-Passante-Persico95, Salam10, Passante18}. A remarkable property of these effects is their dependence on the geometry of the system, as well as on the magneto-dielectric properties of the  materials \cite{Bordag-Klimchitskaya09, Spagnolo-Dalvit07, Palacino-Passante17, Armata-Butera17}, and it has been shown  that they can be controlled through a structured environment (cavities, waveguides, photonic crystals) \cite{Incardone-Fukuta14, Notararigo-Passante18, Palacino-Passante17,John-Quang94}.  The effect of a static external field on the dispersion interactions between two atoms has been also considered \cite{Fiscelli-Rizzuto18, Fiscelli-Rizzuto20, Abrantes-Pessanha21, Karimpour-Reza22a, Karimpour-Reza22b}. 

These effects have been also investigated in dynamical situations, for example when atoms or macroscopic objects are set in motion or when some physical parameter of the system is modulated in time \cite{Dodonov20,  Reiche-Intravaia22, Antezza-Braggio14}. For example,  real photons can be produced from the vacuum by a non adiabatic change of some boundary in the system (Dynamical Casimir Effect) or when an atom is subjected to an oscillatory motion in vacuum (microscopic Dynamical Casimir Effect) \cite{Melo-Impens18, Lo-Law18}.      
Also,  a {\em frictional force} may arise when two or more objects move relative to each other in vacuum at a constant velocity, an effect usually known as quantum friction \cite{Reiche-Intravaia22, Belen-Fosco19}. 
The effects of uniform acceleration of neutral atoms in vacuum and their relation with the Unruh effects have been also studied \cite{Crispino-Higuchi08, Soda-Sudhir22}.
Dynamical environments can also modify radiative properties of atoms or molecules nearby, under adiabatic conditions \cite{Glaetze-Hammerer10, Ferreri-Domina19, Reina-Domina21, Calajo-Rizzuto17}.  

A conceptually different situation also giving a dynamical (i.e. time-dependent) effect, is when some physical parameter in the system is suddenly changed, so that the system evolves starting from a non-equilibrium situation.  
For example,  a sudden change of some atomic parameter, such as the atomic transition frequency between two levels or the electric dipole moment of one atom, gives rise to time-dependent effects, such as the dynamical Lamb shift, the dynamical Casimir-Polder force between two atoms \cite{Passante-Persico02} or the dynamical Casimir-Polder force between an atom and a conducting mirror \cite{Vasile-Passante08, Messina-Vasile10, Haakh-Henkel14}. From the physical point of view, these dynamical effects are essentially related to a change of the virtual photon cloud around the atom due to a rapid change of some physical characteristic of the atom by an external action \cite{Passante-Vinci96, Rizzuto-Passante04, Armata-Vasile16}.  It should be stressed that no real photons creation is taken into account in this case: our time-dependent effects arise from the non-equilibrium dynamics of the system, that evolves unitarily under the  new Hamiltonian.
We note that non-equilibrium conditions may manifest in many realistic physical situations, for example when two or more interacting bodies in a system can be at different temperatures from each other or from the temperature of the environment \cite{Noto-Messina14}. 

The main aim of this paper is to investigate the time-dependent atom-wall Casimir-Polder interaction energy between a ground-state atom and a plane mirror, when the system evolves from a nonequilibrium initial state,
after a sudden change of the atomic position with respect to the mirror.  Specifically, we consider an atom prepared in its dressed ground state, in the presence of a reflecting plane mirror.  This state is an eigenstate of the total atom-field interacting Hamiltonian.  We then suppose that at $t=0$ the atomic position (with respect to the mirror) is abruptly changed by some external action (quench of the atomic position). This change brings the system to a non equilibrium situation since the initial (dressed) ground state of the system  is no longer an eigenstate of the new Hamiltonian, and the system starts to evolve in time. We investigate the time-dependent atom-wall Casimir-Polder force during the dynamical evolution of the system, leading it to the new equilibrium configuration.
We first solve the Heisenberg equations of motion of the atom-field interacting system by an iterative procedure,  and then evaluate explicitly the time-dependent atom-wall Casimir–Polder interaction energy during the atomic self-dressing process. We find that, although the atom is in its ground state, the dynamical Casimir-Polder force exhibits oscillations in time, and can be attractive or repulsive depending on the atomic position and time. This is at variance with the usual (static) Casimir-Polder force between a ground-state atom and a mirror, where the force is always attractive. In the limit of long times it settles to the usual attractive Casimir-Polder interaction between a ground-state atom and a perfectly reflecting plate. 
We also discuss the time-dependent expectation values of global observables, such as  the atomic and field Hamiltonians, on the partially dressed state and their asymptotic behavior, and discuss some aspects of the dynamical process bringing the interaction energy to the new equilibrium situation.  
Our results show that new effects can appear under non equilibrium situations and that the Casimir-Polder force  can exhibit new features in dynamical situations. 

\section{\label{sec:level2} Dynamical atom-wall Casimir-Polder interaction energy}

\begin{figure}[H]
\centering
\includegraphics[width=8.5 cm]{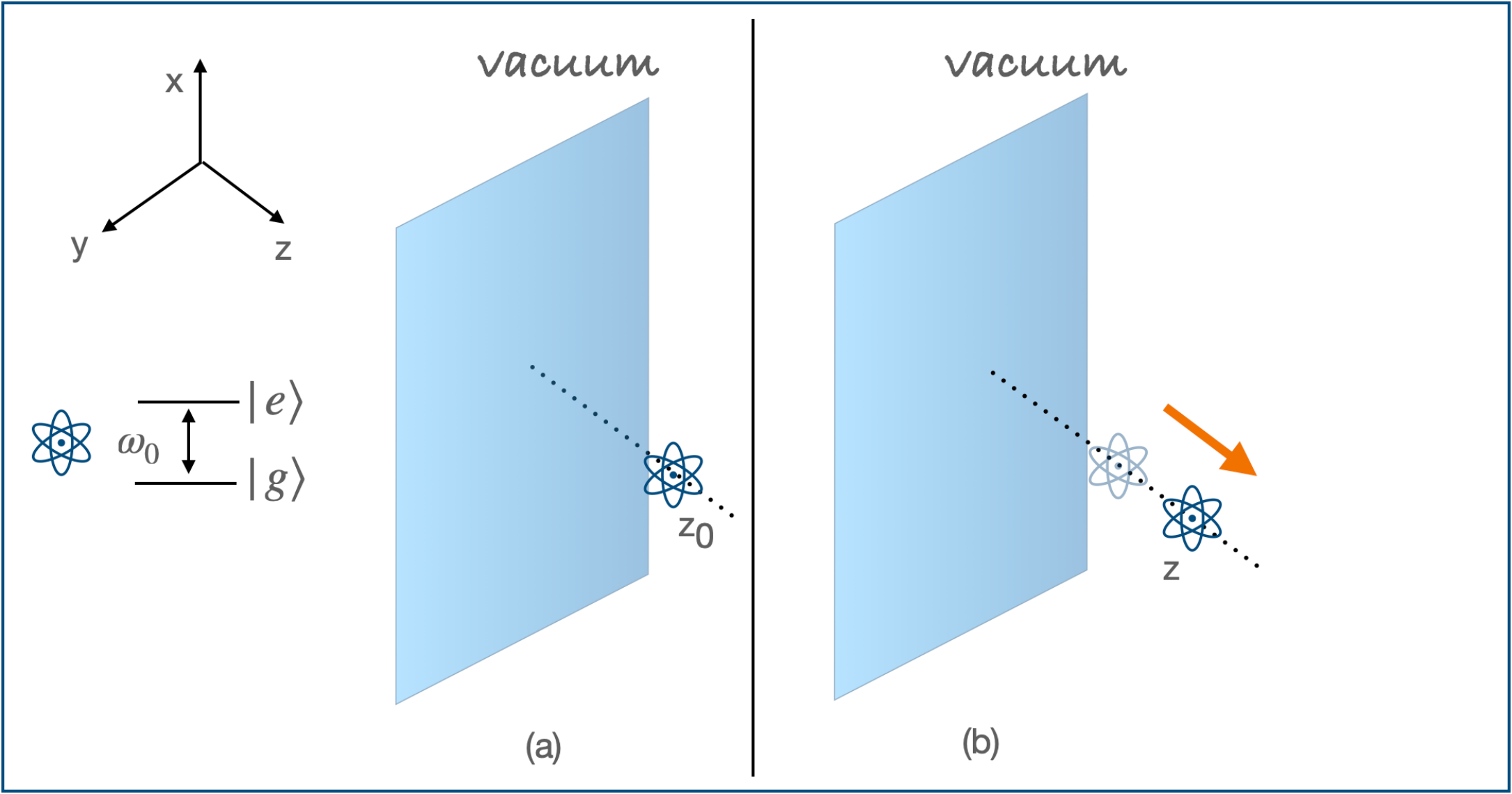}
\caption{Pictorial description of the system. $(a)$ The atom, modeled as two-level system, is located at a distance $\vert\br_0\vert=z_0$ from the mirror at $z=0$.  $(b)$ At $t=0$ the atom is subjected to a rapid change of its position, from $\vert\br_0\vert= z_0$ to the new position $\vert\br\vert=z$ with respect to the mirror.} 
\label{Fig1}
\end{figure}

Our system consists of a neutral atom, modeled as a two-level system, placed at a distance $\vert {\br_0}\vert=z_0$ from an infinite perfectly conducting plate, and interacting with the quantum electromagnetic field in the vacuum state (see Fig. (\ref{Fig1}(a))). 

We adopt the Hamiltonian in the multipolar coupling scheme within the dipole approximation,  in the Coulomb gauge; the interaction Hamiltonian is then given by $H_I = -\bm\cdot{\bE}({\bf x})$, where $\bm$ is the atomic dipole moment operator and ${\bf E}({\bf x})$ is the transverse displacement field at the atomic position ${\bf x}$. 

The total Hamiltonian describing the atom-field interacting system, before the change of the atomic position, is given by \cite{Power-Zienau59}
\begin{eqnarray}
\label{eq:2.1}
& &\tilde{H}=\hbar\w0S_z+\sum_{\bk j}\hbar\wk\akjd\akj+\tilde{V}_1+\lambda \tilde{V}_2,\\
& & \tilde{V}_1=\sum_{\bk j}\left(\epsilon_{\bk j}(\br_0)\akj S_{+} + \epsilon^{\star}_{\bk j}(\br_0)\akjd S_{-}\right),
\,\,\, \tilde{V}_2=\sum_{\bk j}\left(\epsilon_{\bk j}(\br_0)\akjd S_{+} + \epsilon^{\star}_{\bk j}(\br_0)\akj S_{-}\right),\nonumber
\end{eqnarray}  
\noindent where $\akj$ and $\akjd$ are the usual bosonic annihilation and creation field operators, $S_+$, $S_-$ and $S_z$ are the atomic pseudospin 
operators, $\omega_0 $ is the atomic transition frequency. Also, $\epsilon_{\bk j}(\br_0)=-i\sqrt{\frac{2\pi\hbar c}{V}}\sum_{\bk j} \sqrt{k}\left[\bm\cdot {\bf f}_{\bk j}(\br_0)\right]$ is the coupling constant and ${\bf f}_{\bk j}(\br)$ are the usual field mode functions satisfying the boundary conditions on the perfectly conducting walls of a cubic cavity \cite{Armata-Vasile16}.
Finally, $\lambda$ is a parameter that can assume the values $\lambda=0,-1$, so that for $\lambda=0$ we obtain the Hamiltonian in the Rotating Wave Approximation (RWA).

We assume the atom initially prepared in its dressed ground state, that is an eigenstate of total Hamiltonian \eqref{eq:2.1}; at  first-order in the atom-field coupling we have,
\begin{eqnarray}
\label{eq:2.3}
\tilde{\ket{g}}_d=\ket{g,\{0_{\bk j}\}}-\lambda\sum_{\bk j}\frac{\epsilon_{\bk j}(\br_0)}{\hbar (\wk+\w0)} \ket{e, 1_{\bk j}}\, .
\end{eqnarray}

We now suppose that at time $t=0$ the atomic position is abruptly changed by some external action, from $\br_0$ to the final position $\br$ (see figure (\ref{Fig1}(b))).  
This is equivalent to a rapid change of the atom-field coupling that {\em instantaneously} brings the system to a dynamical non-equilibrium configuration.
Our assumption of a sudden change of the atomic position is valid if this change occurs in a time $t$ such that $t\ll 2r_0/c$ (that is the time taken by a virtual photon emitted by the atom to go back after reflection on the mirror) and $t\ll\omega_0^{-1}$. 
For an atom-plate distance of the order of $10^{-7}\, \mbox{m}$, our non-adiabatic approximation requires times of the order of $10^{-16}\, \mbox{s}$, which is also consistent with the requirement $t\ll\omega_0^{-1}$ for a typical atomic transition frequency $\omega_0\sim 10^{15}\, \mbox{s}^{-1}$.

As a consequence of this change, the ground state $\tilde{\ket{g}}_d$ in \eqref{eq:2.3} it is not longer an eigenstate of the new Hamiltonian which is now
\begin{eqnarray}
\label{eq:2.4}
H&=&\hbar\w0 S_z+\sum_{\bk j}\hbar\wk\akjd\akj+ \sum_{\bk j}\left(\epsilon_{\bk j}(\br)\akj S_{+} + \epsilon^{\star}_{\bk j}(\br)\akjd S_{-}\right) \nonumber\\
&\ &+ \lambda  \sum_{\bk j}\left(\epsilon_{\bk j}(\br)\akjd S_{+} + \epsilon^{\star}_{\bk j}(\br)\akj S_{-}\right)\, ,
\end{eqnarray}  
\noindent and the system will evolve in time starting from the initial partially dressed state $\ket{\psi}_p$, under the Hamiltonian  \eqref{eq:2.4} (note that the only difference between the new Hamiltonian \eqref{eq:2.4} and the old one in \eqref{eq:2.1} is the dependence on the atomic position in the coupling constant, which is now $\epsilon_{\bk j}(\br)$ and not $\epsilon_{\bk j}(\br_0)$). It is worth noting that the bare ground-state $\ket{g,\{0_{\bk j}\}}$ is an exact eigenstate of the total Hamiltonian in the RWA ($\lambda=0$), and therefore it does not evolve.   
 We stress that in our approximation, no time dependence is included in the system's Hamiltonian, the atom being supposed at rest and locally interacting with the vacuum field fluctuations at the new (static) atomic position. Therefore, no real photons creation is involved in this case, time-dependent effects only resulting from the non-equilibrium dynamics of the system that evolves unitarily under the new Hamiltonian \eqref{eq:2.4}.  Time-dependent effects are only due to the atomic self-dressing process after the rapid perturbation of the virtual photons cloud surrounding the atom. Also, under these assumptions, we can neglect the contribution from the R{\" o}ntgen term in the Hamiltonian, which is of the order of $v/c$, where $v$ is the velocity of the atom. 
To evaluate the time-dependent Casimir-Polder interaction,  
we follow the same procedure adopted in \cite{Vasile-Passante08, Messina-Vasile10, Haakh-Henkel14, Armata-Vasile16} where the second-order interaction energy is obtained as half of the
average value of the interaction Hamiltonian $H_I^{(2)}(t)$ on the initial state of the system (in our case, the partially dressed state at $t=0^+$ immediately after the change of the atomic position), and taking only terms depending on the atom-plate distance,
\begin{eqnarray}
\label{eq:2.5}
\Delta E^{(2)}(t)= \frac 1 2\,\,\, _p\!\bra{\psi}H_I^{(2)}(t)\ket{\psi}_p, 
\end{eqnarray}

\noindent where $\ket{\psi}_p$ is the partially dressed state of the system, and  $H_I^{(2)}(t)$  is the (second-order) interaction Hamiltonian in the Heisenberg picture, obtained by solving iteratively the Heisenberg equations of motion for field and atomic operators.  
Equation \eqref{eq:2.5} is a generalization to time-dependent situations of the well-known expression of the interaction energy obtained by second-order stationary perturbation theory \cite{Compagno-Passante83}, $\Delta E=\frac1 2\,\,\, {_d\!\bra{g}} H_I\ket{g}\!_d ,$
where $\ket\psi_d$ is the second-order dressed state of the system, and $H_I$ is the interaction Hamiltonian in the Schr\"{o}dinger representation.
We note that, since the time evolution of the system is unitary, the total energy of the system is conserved during the dynamical self-dressing process. However, this energy does not coincide with the total energy of the (second-order) dressed state $\ket{g}_d$ of the Hamiltonian \eqref{eq:2.4}, that is $_p\!\bra{\psi} H^{(2)}(t)\ket{\psi}_p\not=_d\!\!\bra{g} H \ket{g}_d$. 
This is because dressed and partially dressed states have different energy (we shall discuss this relevant point below). 
Nevertheless, as we now show, the average value of the interaction Hamiltonian $H_I^{(2)}(t)$ on the partially dressed state $\ket\psi_p$, settles to its stationary value $_d\!\bra{g} H_I \ket{g}_d$ in the limit of large times ($t\rightarrow\infty$), and gives the usual atom-plate Casimir-Polder interaction energy in the static case.  
 
We first calculate the expressions of atomic and field operators in the Heisenberg representation. At the lowest significant order, we obtain
\begin{eqnarray}
\label{eq:2.7}
S_{+}(t)=e^{i\w0 t}S_{+}(0)+\frac{2i}{\hbar}S_z(0)e^{i\w0 t}\sum_{\bk j}\epsilon_{\bk j}(\br)[\akjd (0)G(\wk-\w0,t)+\lambda\akj(0)G^{\star}(\wk+\w0,t)] \, ,\\
\label{eq:2.8}
\akj(t)=e^{-i\wk t} \akj (0)+\frac{i}{\hbar}e^{-i\wk t}\epsilon_{\bk j}(\br)\left(S_{-}(0)G(\wk-\w0,t)-\lambda S_{+}G(\wk+\w0,t)\right)\, , 
\end{eqnarray}
\noindent where we have introduced the function
\begin{equation}
G(x,t)=\int_0^{t}e^{i x t'}dt' = \frac{e^{i x t}-1}{ix}\, . 
\label{eq:2.9}
\end{equation}

Substituting equations \eqref{eq:2.7} and \eqref{eq:2.8} into the expression of the interaction Hamiltonian  $V_1+\lambda V_2$ (see Eq. \eqref{eq:2.4}), and  
taking the average value on the partial dressed state  $\ket \psi_p$, we obtain (up to second order in the coupling): 
\begin{eqnarray}
& & {\cal{E}}(t)_I= _p\!\bra{\psi}H_I^{(2)}(t)\ket{\psi}\!_p =-\lambda^2\frac{4\pi }{V}\sum_{\bk j}\left( \bm\cdot{\bf f}_{\bk j}(\br)\right)^2\frac{\wk}{\wk+\w0}\left (1-\cos(\w0+\wk)t\right)\nonumber\\
& & -\lambda^2\frac{4\pi }{V}\sum_{\bk j}\left(\bm\cdot{\bf f}_{\bk j}(\br)\right) \left(\bm\cdot{\bf f}_{\bk j}(\br_0)\right)\frac{\wk}{\wk+\w0}\cos(\w0+\wk)t\, .
\label{eq:2.11}
\end{eqnarray} 

The presence of the parameter $\lambda$ in the equation above clearly shows that the time-dependent interaction energy is related to the virtual photons that are emitted and reabsorbed by the atom during the self-dressing process of the system. In other words, energy conserving processes arising from the rotating terms of the interaction Hamiltonian and involving real photons, do not play any role at the order considered. 
     
In the continuum limit, using $\sum_{\bk}\rightarrow V/(2\pi)^3 \int d^3{\bf k}$, after some algebra, we get
\begin{eqnarray}
\label{eq:2.12}
&\ &{\cal{E}}(t)_I = \Delta E^{(2)}_{f.s.}(t)+\Delta E^{(2)}_{b.}(t)\, ,
\end{eqnarray}
\noindent where
\begin{eqnarray}
\label{eq:2.12a}
\Delta E^{(2)}_{f.s.}(t) &=& -2\frac{\bm_{\ell}\bm_m}{\pi}\Biggl\{\frac 2 3 \delta_{\ell m}\int_0^{\infty} dk k^3 \frac{1-\cos\left(c t(k+k_0\right)}{k+k_0}\nonumber\\
&\ &+ F_{\ell m}^{R'}\frac{1}{R'}\int_0^{\infty}d k \frac{\sin kR'}{k+k_0}\cos\left(c t(k+k_0)\right)\Biggr\}\, ,
\end{eqnarray}
\noindent is the free-space contribution to the time-dependent interaction energy of the partially dressed atom ($R'=\vert \br-\br_0\vert$, and $F_{\ell m}^x=(-\delta_{\ell m}\nabla^2+\nabla_{\ell}\nabla_{m})^x$ is a differential operator acting on the variable $x$, and we used the Einstein convention for repeated indices), and
\begin{eqnarray}
\label{eq:2.12b}
\Delta E^{(2)}_{b.}(t) &=& \frac{\bm_{\ell}\bm_m}{\pi}\Biggl\{ \sigma_{\ell p}\biggl[F_{pm}^{\bar{R}}\frac{1}{\bar{R}}\int_0^{\infty}d k \frac{\sin k\bar{R}}{k+k_0}(1-\cos\left(c t(k+k_0\right))\nonumber\\
&\ &+ F_{pm}^{\bar{R'}}\frac{1}{\bar{R'}}\int_0^{\infty}d k \frac{\sin k\bar{R'}}{k+k_0}\cos\left(c t(k+k_0\right)\biggr]\Biggr\}\, ,
\end{eqnarray}
\noindent is the boundary-dependent  contribution to the interaction energy. Here, $\bar{R}=\vert \br-\sigma\br\vert$ is the distance between the atomic position $\br$ and its image through the mirror, and $\bar{R}'=\vert \br-\sigma\br_0\vert$, where $\sigma=\text{diag}(1,1,-1)$ is the reflection matrix . 

The expressions above give the time-dependent interaction energy during the self-dressing process of the atom after the abrupt change of atom's position. In particular, the free-space term \eqref{eq:2.12a} can be interpreted as a time-dependent Lamb-shift due to the nonadiabatic perturbation of the virtual photons cloud surrounding the atom.    
On the other hand, the boundary-dependent contribution \eqref{eq:2.12b} gives the time-dependent atom-wall Casimir-Polder interaction energy. 

\begin{figure}[H]
\centering
\includegraphics[width=8.0 cm]{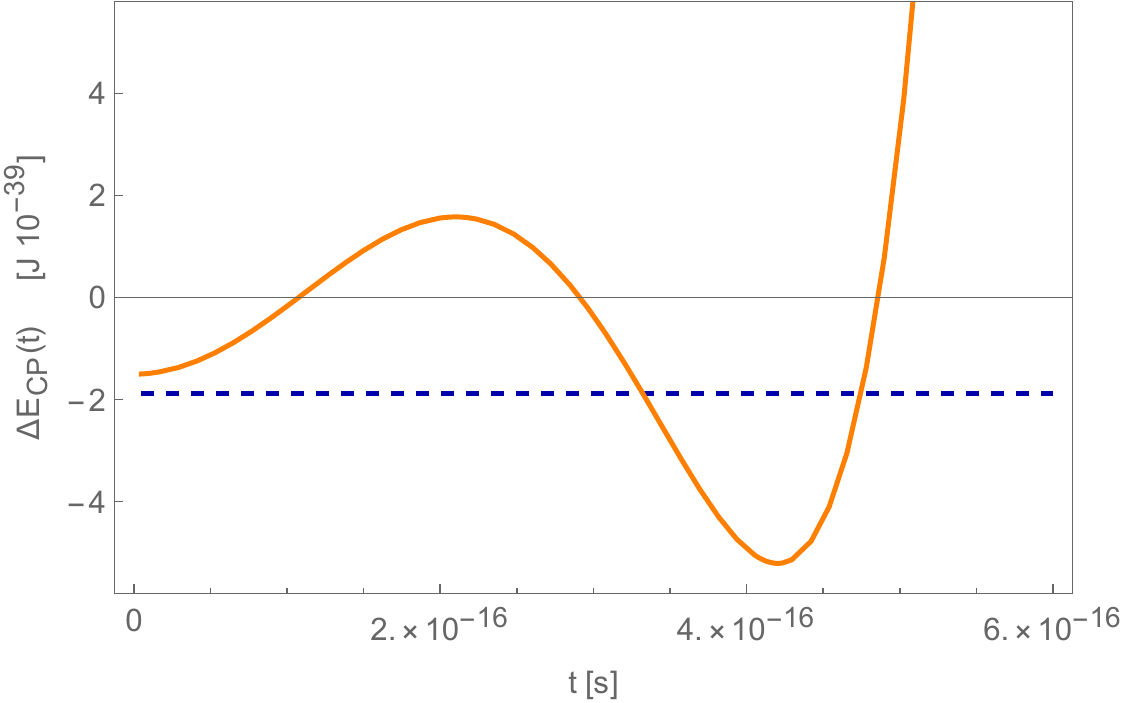}
\caption{Dynamical Casimir-Polder interaction energy for $t<\bar{R}/c$. The orange (solid) line corresponds to the case of an initial partially dressed state, while the blue (dashed) line refers to the static CP force after the quench. The figure shows that the value of the interaction energy at $t=0^+$ is non vanishing, as a consequence of the assumption of an initial partially dressed state. The values used are $\mu=6.31\times10^{-30}$ \mbox{C m}, $k_0=5.0\times10^7$ \mbox{m$^{-1}$}, $r_0=1.001\times10^{-7}$ \mbox{m}, $r=1.0\times10^{-7}$ \mbox{m} } \label{Plot1}.
\end{figure}

\begin{figure}[H]
\centering
\includegraphics[width=8.0 cm]{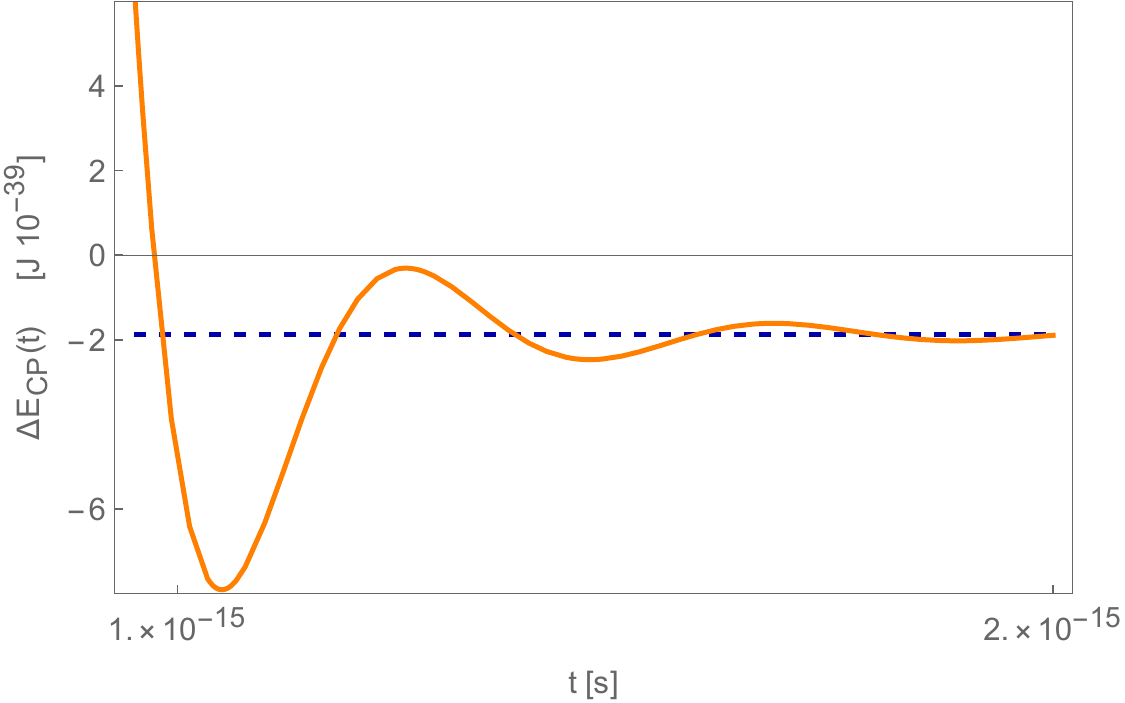}
\caption{Dynamical Casimir-Polder interaction energy for $t>\bar{R}/c$. The orange (solid) line corresponds to the time-dependent interaction energy for $t > \bar{R}/c$, while the blue (dashed) line refers to the static interaction energy in the asymptotic limit. $\mu=6.31\times10^{-30}$ \mbox{C m}, $k_0=5.0\times10^{7}$ \mbox{m$^{-1}$}, $r_0=1.001\times10^{-7}$ \mbox{m}, $r=1.0\times10^{-7}$ \mbox{m} }.
\label{Plot2}
\end{figure}

We now focus on this last term with the aim to discuss some features of this dynamical interaction energy.   
Using equation \eqref{eq:2.12b} in  \eqref{eq:2.5}, for an atom with an isotropic electric dipole moment ($\vert \bm_x\vert^2=\vert \bm_y\vert^2=\vert \bm_z\vert^2=\vert \bm\vert^2/3$),  we obtain 
\begin{eqnarray}
\label{eq:2.13}
 \Delta E^{(2)}_{CP}(t)=\frac 12 \Delta E^{(2)}_{b.}(t)&=&
-\frac{2\mu^2}{3\pi} \lim_{q\rightarrow 1} D_q \Biggl\{\frac{1}{\bar{R}}\int_0^{\infty} d\chi \frac{\sin(q\chi)}{\chi+\chi_0}\left(1-\cos(a(\chi+\chi_0))\right)\Biggr.\nonumber\\
& &\Biggl. - \frac{1}{\bar{R'}}\int_0^{\infty}\frac{\sin(q\bar{\chi})}{\bar{\chi}+\bar{\chi}_0}\left(cos(\bar{a}(\bar{\chi}+\bar{\chi}_0))\right)\Biggr\}
\end{eqnarray}
where we have defined the differential operator $D_q=2-2\partial/\partial q+\partial^2/\partial q^2$, $a=\frac{ct}{\bar{R}}$, and $\bar{a}=\frac{ct}{\bar{R}'}$. Also,  $\chi=k\bar{R}$, $\chi_0=k_0\bar{R}$,  $\bar{\chi}=k\bar{R}'$ and $\bar{\chi}_0=k_0\bar{R}'$. 
The integrals in \eqref{eq:2.13} can be explicitly calculated 
for $a<1$ (i.e. $t<\bar{R}/c$) and $a>1$ (i.e. $t>\bar{R}/c$) (note that the integrals diverge on the light cone $c t=\bar{R}$; the physical origin of this divergence can be ascribed to the dipole approximation and the radiation reaction field \cite{Vasile-Passante08, Messina-Vasile10}).  
We do not give here the explicit expression of the time-dependent Casimir-Polder interaction energy, because the relevant features of the atom-plate interaction can be extrapolated from figures \ref{Plot1} and \ref{Plot2}. These figures show  the time evolution of the Casimir-Polder interaction energy, near the light-cone, for $ct<\bar{R}$ and $ct>\bar{R}'$ (we have supposed that the atom's position is rapidly changed of a distance of the order $10^{-10}$ \mbox{m} from the initial position $z_0=10^{-7}\, \mbox{m}$). Similarly to the case discussed in \cite{Messina-Vasile10}, the Casimir-Polder interaction energy oscillates in time, switching in time from an attractive to a repulsive force. In the asymptotic limit ($t\rightarrow\infty$), the interaction energy settles to its stationary value. The oscillations of the interaction energy and its strong increase around the round-trip time $t=\bar{R}/c$ (typically of the order of $10^{-16}\div 10^{-15}$ \mbox{s},) suggest the possibility of observing these effects in laboratory. Also, due to the assumption of an initial partially dressed state, the value of the interaction energy at $t=0^+$ is not zero; it represents a part of the energy stored in the system immediately after the change the atom's position. 
 To make this point clearer, it is convenient to reconsider the interaction energy of the partially dressed state given in \eqref{eq:2.11}. With a simple algebraic manipulation, this expression can be written as follows
\begin{eqnarray}
& &{\cal{E}}_I(t) =\,_p\!\bra{\psi}H_I^{(2)}(t)\ket{\psi}\!_p =-\frac{4\pi }{V}\sum_{\bk j}\left( \bm\cdot{\bf f}_{\bk j}(\br)\right)^2\frac{\wk}{\wk+\w0}\nonumber\\
& & +\frac{4\pi }{V}\sum_{\bk j}\bm\cdot{\bf f}_{\bk j}(\br)\left (\bm\cdot{\bf f}_{\bk j}(\br)-\bm\cdot{\bf f}_{\bk j}(\br_0)\right)\frac{\wk}{\wk+\w0}\cos(\w0+\wk)t\, ,
\label{eq:2.17}
\end{eqnarray} 

\noindent that clearly shows that the interaction energy at the initial time is different from zero and depends on $\br_0$ and $\br$. Thus, it keeps memory of the atom's position before and after its sudden change. On the other hand, in the limit of large times, the average value of the interaction energy settles to its static value in the new equilibrium configuration of the system, that is
\begin{eqnarray}
\label{eq:2.18}
\Delta E^{(static)}_I = \frac 12\,\,\,\lim_{t\rightarrow\infty}\,\, _p\!\bra{\psi}H_I^{(2)}(t)\ket{\psi}\!_p =\frac 12\,\,\,_d\,\!\!\bra{g^{(2)}}H_I\ket{g^{(2)}}\!_d  
\end{eqnarray}   

\noindent where $\ket{g^{(2)}}\!_d$ is the fully (second-order) dressed state of the interacting system (with the atom located at position $\br$).


Thus, we find that the system evolves in time and the interaction energy reaches a new equilibrium condition in the long-time limit. 
It is  
also interesting to evaluate the time-dependent average values of global observables, such as the field and atomic Hamiltonians. 
To do that, we first calculate the second-order dressed ground state of the total Hamiltonian \eqref{eq:2.1}. We get,

\begin{eqnarray}
\label{eq:2.19}
\ket{\tilde{g}^{(2)}}_d&=&\left(1-\frac{\lambda^2}{2} \sum_{\bk j}\frac{\vert \epsilon_{\bk j}(\br_0)\vert^2}{\hbar^2(\wk+\wk)^2}\right)\ket{g,\{0_{\bk j}\}}-\lambda\sum_{\bk j}\frac{\epsilon_{\bk j}(\br_0)}{\hbar (\wk+\w0)} \ket{e, 1_{\bk j}}\nonumber\\
&\ &-\lambda\sum_{\bk j}\sum_{\bk' j'}\frac{\epsilon_{\bk j}(\br_0)\epsilon_{\bk' j'}(\br_0)}{\hbar^2(\w0+\omega_{k'})(\w0+\wk)}\vert g, 1_{\bk j},1_{\bk' j'}\rangle\, .
\end{eqnarray}
As before, this state is a partially dressed state after the {\em nonadiabatic} change of the atomic position and evolves dynamically under the Hamiltonian \eqref{eq:2.4}. The time-dependent average values of atomic and field Hamiltonian on the partially dressed state \eqref{eq:2.19} are easily evaluated, obtaining

\begin{eqnarray}
\label{eq:2.20}
{\cal{E}}_A(t)&=&\,_p\langle  \psi^{(2)}\vert H_A^{(2)}(t)\vert  \psi^{(2)}\rangle_p= \frac{2\pi}{V}\sum_{{\bf k}j}\left({\boldsymbol\mu}\cdot{\bf f}_{{\bf k}j}({\bf r}_0)\right)^2 \frac{\omega_0\omega_k}{(\omega_0+\omega_k)^2}\nonumber\\
&\ &+\frac{4\pi}{V}\sum_{{\bf k}j}\left({\boldsymbol\mu}\cdot{\bf f}_{{\bf k}j}(\bf r)\right)\left({\boldsymbol\mu}\cdot{\bf f}_{{\bf k}j}(\br)-{\boldsymbol\mu}\cdot{\bf f}_{{\bf k}j}({\bf r}_0)\right)\frac{\omega_0\omega_k}{(\omega_0+\omega_k)^2}\left(1-\cos(\omega_0+\omega_k)t\right),
\end{eqnarray}
and 
\begin{eqnarray}
\label{eq:2.21}
{\cal{E}}_F(t)&=&\, _p\langle \psi^{(2)}\vert H_F^{(2)}(t)\vert \psi^{(2)}\rangle_p= \frac{2\pi}{V}\sum_{{\bf k}j}\left({\boldsymbol\mu}\cdot{\bf f}_{{\bf k}j}({\bf r}_0)\right)^2 \frac{\omega_k^2}{(\omega_0+\omega_k)^2}\nonumber\\
&\ &+\frac{4\pi}{V}\sum_{{\bf k}j}\left({\boldsymbol\mu}\cdot{\bf f}_{{\bf k}j}(\bf r)\right)\left({\boldsymbol\mu}\cdot{\bf f}_{{\bf k}j}(\br)-{\boldsymbol\mu}\cdot{\bf f}_{{\bf k}j}({\bf r}_0)\right)\frac{\omega_k^2}{(\omega_0+\omega_k)^2}\left(1-\cos(\omega_0+\omega_k)t\right).
\end{eqnarray}
\noindent In the asymptotic limit ($t\rightarrow\infty$), we have
\begin{eqnarray}
\label{eq:2.22}
{\cal{E}}_A(t)\rightarrow \frac{2\pi}{V}\sum_{{\bf k}j} \frac{\omega_0\omega_k\left({\boldsymbol\mu}\cdot{\bf f}_{{\bf k}j}({\bf r}_0)\right)^2}{(\omega_0+\omega_k)^2}+\frac{4\pi}{V}\sum_{{\bf k}j}\frac{\omega_0\omega_k\left({\boldsymbol\mu}\cdot{\bf f}_{{\bf k}j}(\bf r)\right)}{(\omega_0+\omega_k)^2}\left({\boldsymbol\mu}\cdot{\bf f}_{{\bf k}j}(\br)-{\boldsymbol\mu}\cdot{\bf f}_{{\bk}j}({\bf r}_0)\right)\, ,
\end{eqnarray}
and
\begin{eqnarray}
\label{eq:2.24}
{\cal{E}}_F(t)\rightarrow \frac{2\pi}{V}\sum_{{\bf k}j} \frac{\omega_k^2\left({\boldsymbol\mu}\cdot{\bf f}_{{\bf k}j}({\bf r}_0)\right)^2}{(\omega_0+\omega_k)^2}+\frac{4\pi}{V}\sum_{{\bf k}j}\frac{\omega_k^2\left({\boldsymbol\mu}\cdot{\bf f}_{{\bf k}j}(\bf r)\right)}{(\omega_0+\omega_k)^2}\left({\boldsymbol\mu}\cdot{\bf f}_{{\bf k}j}(\br)-{\boldsymbol\mu}\cdot{\bf f}_{{\bk}j}({\bf r}_0)\right)\, .
\end{eqnarray}

A comparison with the analogous quantities obtained in the static case shows that, in the limit of large times, the time-dependent atomic and field energies does not settle to their stationary values on the fully dressed state of the system; the additional terms in \eqref{eq:2.22} and \eqref{eq:2.24} are related to the energy initially stored in the  field and in the atom, and to the {\em work} done to bring the system in out-of-equilibrium configuration. In fact, we have
\begin{eqnarray}
\label{eq:2.23}
&\ &\lim_{t\rightarrow\infty}\,{\cal{E}}_A(t)\not={\cal{E}}_A^{stat.}= _d\!\bra{g^{(2)}} H_A\ket{g^{(2)}}_d=\frac{2\pi}{V}\sum_{{\bf k}j}\left({\boldsymbol\mu}\cdot{\bf f}_{{\bf k}j}({\bf r})\right)^2 \frac{\omega_0\omega_k}{(\omega_0+\omega_k)^2}\, ,\\
\label{eq:2.25}
&\ &\lim_{t\rightarrow\infty}\, {\cal{E}}_F(t)\not={\cal{E}}_F^{stat.}= _d\!\bra{g^{(2)}} H_F\ket{g^{(2)}}_d=\frac{2\pi}{V}\sum_{{\bf k}j}\left({\boldsymbol\mu}\cdot{\bf f}_{{\bf k}j}({\bf r})\right)^2 \frac{\omega_k^2}{(\omega_0+\omega_k)^2}\, .
\end{eqnarray} 

This is different from the case where the system evolves from an initial bare state, where in the asymptotic limit both dynamical energies stored in the field and in the atom tend to twice their stationary values \cite{Armata-Vasile16, Armata-Kim17}. On the other hand, at large times the energy stored in the interaction Hamiltonian coincides with its stationary value (see Equation\eqref{eq:2.18}). It should be also stressed that since the time evolution of the system is unitary, the average value of the total Hamiltonian does not change during the time evolution of the system, and it is

\begin{eqnarray}
\label{eq:2.26}
&\ &{\cal{E}}^{part.}_{tot}=\, _p\!\langle  \psi^{(2)}\vert H^{(2)}(t)\vert  \psi^{(2)}\rangle_p={\cal{E}}_A+{\cal{E}}_F+{\cal{E}}_I =-\frac{2\pi}{V}\sum_{{\bf k}j}\left({\boldsymbol\mu}\cdot{\bf f}_{{\bf k}j}({\bf r}_0)\right)^2 \frac{\omega_k}{\omega_0+\omega_k}\nonumber\\
&\ &+\frac{4\pi}{V}\sum_{{\bf k}j}\left({\boldsymbol\mu}\cdot{\bf f}_{{\bf k}j}(\bf r)\right)\left({\boldsymbol\mu}\cdot{\bf f}_{{\bf k}j}(\br_0)\right)\frac{\omega_k}{\omega_0+\omega_k}-\frac{4\pi}{V}\sum_{{\bf k}j}\left({\boldsymbol\mu}\cdot{\bf f}_{{\bf k}j}({\bf r})\right)^2 \frac{\omega_k}{\omega_0+\omega_k}\, .
\end{eqnarray}
  
This is different from what obtained in a completely static situation, that is
\begin{eqnarray}
\label{eq:2.27}
&\ &{\cal{E}}^{stat.}_{tot}=\, _d\!\bra{g^{(2)}} H_{tot}\ket{g^{(2)}}_d=-\frac{2\pi}{V}\sum_{{\bf k}j}\left({\boldsymbol\mu}\cdot{\bf f}_{{\bf k}j}({\bf r})\right)^2 \frac{\omega_k}{\omega_0+\omega_k}\, .
\end{eqnarray}

\noindent Comparison of \eqref{eq:2.26} and \eqref{eq:2.27} shows that for $t\rightarrow\infty$ the total energy, similarly to the atomic and field contributions, does not settle to the value obtained in the stationary case. 

\section{\label{sec:level3} Conclusions}
In this paper, we have investigated the dynamical Casimir-Polder interaction energy between an atom and a perfect reflecting plane during the time evolution of the system after a sudden change of the atomic position, that brings the system to a nonequilibrium situation. We have discussed the dynamical processes leading the interaction energy to the new equilibrium configuration. We have shown that the time-dependent Casimir-Polder interaction energy oscillates in time, changing from attractive to repulsive, and that in the limit of large times it settles to its stationary value. Finally, we have investigated some interesting aspects of the dynamical atomic self-dressing process by investigating the time dependence of the field and atomic Hamiltonians, and discussed the different behaviour of {\em local} and {\em global} observables in the dynamical evolution of the system. 

\ack
R.~P. and L.~R. acknowledge partial financial support from the FFR2021 grant from the University of Palermo, Italy.

\section*{References}
\bibliography{biblio1.bib}

\providecommand{\newblock}{}
\begin{thebibliography}{10}
\expandafter\ifx\csname url\endcsname\relax
  \def\url#1{{\tt #1}}\fi
\expandafter\ifx\csname urlprefix\endcsname\relax\def\urlprefix{URL }\fi
\providecommand{\eprint}[2][]{\url{#2}}

\bibitem{Lamb-Retherford47}
Lamb W~E and Retherford R~C 1947 {\em Phys. Rev.\/} {\bf 72}(3) 241--243
  \urlprefix\url{https://link.aps.org/doi/10.1103/PhysRev.72.241}

\bibitem{Maclay20}
Maclay G~J 2020 {\em Physics\/} {\bf 2} 105--149 ISSN 2624-8174
  \urlprefix\url{https://www.mdpi.com/2624-8174/2/2/8}

\bibitem{Milonni19}
Milonni P 2019 {\em An Introduction to Quantum Optics and Quantum
  Fluctuations\/} (London, U.K.: Oxford University Press)

\bibitem{Casimir48}
Casimir H~B~G 1948 {\em Proc. Akad. Wet. Amsterdam\/} {\bf 51} 793--795 ISSN
  0370-0348

\bibitem{Casimir-Polder48}
Casimir H~B~G and Polder D 1948 {\em Phys. Rev.\/} {\bf 73}(4) 360--372
  \urlprefix\url{https://link.aps.org/doi/10.1103/PhysRev.73.360}

\bibitem{Compagno-Passante-Persico95}
Compagno G, Passante R and Persico F 1995 {\em Atom-Field Interactions and
  Dressed Atoms\/} (Cambridge, UK: Cambridge Universty Press)

\bibitem{Salam10}
Salam A 2010 {\em Molecular Quantum Electrodynamics\/} (Singapore: McGraw-Hill)

\bibitem{Passante18}
Passante R 2018 {\em Symmetry\/} {\bf 10} 735
  \urlprefix\url{https://www.mdpi.com/2073-8994/10/12/735}

\bibitem{Bordag-Klimchitskaya09}
Bordag M, Klimchitskaya G, Mohideen U and Mostepanenko V 2009 {\em Advances in
  the Casimir Effect\/} (Oxford, UK: Oxford Science Publications)

\bibitem{Spagnolo-Dalvit07}
Spagnolo S, Dalvit D~A~R and Milonni P~W 2007 {\em Phys. Rev. A\/} {\bf 75}(5)
  052117 \urlprefix\url{https://link.aps.org/doi/10.1103/PhysRevA.75.052117}

\bibitem{Palacino-Passante17}
Palacino R, Passante R, Rizzuto L, Barcellona P and Buhmann S~Y 2017 {\em
  Journal of Physics B: Atomic, Molecular and Optical Physics\/} {\bf 50}
  154001 \urlprefix\url{https://doi.org/10.1088/1361-6455/aa75f4}

\bibitem{Armata-Butera17}
Armata F, Butera S, Fiscelli G, Incardone R, Notararigo V, Palacino R, Passante
  R, Rizzuto L and Spagnolo S 2017 {\em Journal of Physics: Conference
  Series\/} {\bf 880} 012064
  \urlprefix\url{https://doi.org/10.1088/1742-6596/880/1/012064}

\bibitem{Incardone-Fukuta14}
Incardone R, Fukuta T, Tanaka S, Petrosky T, Rizzuto L and Passante R 2014 {\em
  Phys. Rev. A\/} {\bf 89}(6) 062117
  \urlprefix\url{http://link.aps.org/doi/10.1103/PhysRevA.89.062117}

\bibitem{Notararigo-Passante18}
Notararigo V, Passante R and Rizzuto L 2018 {\em Scientific Reports\/} {\bf
  8}(1) 5193 \urlprefix\url{https://doi.org/10.1038/s41598-018-23416-0}

\bibitem{John-Quang94}
John S and Quang T 1994 {\em Phys. Rev. A\/} {\bf 50}(2) 1764--1769
  \urlprefix\url{http://link.aps.org/doi/10.1103/PhysRevA.50.1764}

\bibitem{Fiscelli-Rizzuto18}
Fiscelli G, Rizzuto L and Passante R 2018 {\em Phys. Rev. A\/} {\bf 98}(1)
  013849 \urlprefix\url{https://link.aps.org/doi/10.1103/PhysRevA.98.013849}

\bibitem{Fiscelli-Rizzuto20}
Fiscelli G, Rizzuto L and Passante R 2020 {\em Phys. Rev. Lett.\/} {\bf 124}(1)
  013604
  \urlprefix\url{https://link.aps.org/doi/10.1103/PhysRevLett.124.013604}

\bibitem{Abrantes-Pessanha21}
Abrantes P~P, Pessanha V, Farina C and Souza R~d~M~e 2021 {\em Phys. Rev.
  Lett.\/} {\bf 126}(10) 109301
  \urlprefix\url{https://link.aps.org/doi/10.1103/PhysRevLett.126.109301}

\bibitem{Karimpour-Reza22a}
Karimpour M~R, Fedorov D~V and Tkatchenko A 2022 {\em Phys. Rev. Research\/}
  {\bf 4}(1) 013011
  \urlprefix\url{https://link.aps.org/doi/10.1103/PhysRevResearch.4.013011}

\bibitem{Karimpour-Reza22b}
Karimpour M~R, Fedorov D~V and Tkatchenko A 2022 {\em The Journal of Physical
  Chemistry Letters\/} {\bf 13} 2197--2204 pMID: 35231170 (\textit{Preprint}
  \eprint{https://doi.org/10.1021/acs.jpclett.1c04222})
  \urlprefix\url{https://doi.org/10.1021/acs.jpclett.1c04222}

\bibitem{Dodonov20}
Dodonov V 2020 {\em Physics\/} {\bf 2} 67--104 ISSN 2624-8174
  \urlprefix\url{https://www.mdpi.com/2624-8174/2/1/7}

\bibitem{Reiche-Intravaia22}
Reiche D, Intravaia F and Busch K 2022 {\em APL Photonics\/} {\bf 7} 030902
  (\textit{Preprint} \eprint{https://doi.org/10.1063/5.0083067})
  \urlprefix\url{https://doi.org/10.1063/5.0083067}

\bibitem{Antezza-Braggio14}
Antezza M, Braggio C, Carugno G, Noto A, Passante R, Rizzuto L, Ruoso G and
  Spagnolo S 2014 {\em Phys. Rev. Lett.\/} {\bf 113}(2) 023601
  \urlprefix\url{https://link.aps.org/doi/10.1103/PhysRevLett.113.023601}

\bibitem{Melo-Impens18}
Souza R~d~M~e, Impens F~m~c and Neto P~A~M 2018 {\em Phys. Rev. A\/} {\bf
  97}(3) 032514
  \urlprefix\url{https://link.aps.org/doi/10.1103/PhysRevA.97.032514}

\bibitem{Lo-Law18}
Lo L and Law C~K 2018 {\em Phys. Rev. A\/} {\bf 98}(6) 063807
  \urlprefix\url{https://link.aps.org/doi/10.1103/PhysRevA.98.063807}

\bibitem{Belen-Fosco19}
Far\'{\i}as M~B, Fosco C~D, Lombardo F~C and Mazzitelli F~D 2019 {\em Phys.
  Rev. D\/} {\bf 100}(3) 036013
  \urlprefix\url{https://link.aps.org/doi/10.1103/PhysRevD.100.036013}

\bibitem{Crispino-Higuchi08}
Crispino L~C~B, Higuchi A and Matsas G~E~A 2008 {\em Rev. Mod. Phys.\/} {\bf
  80}(3) 787--838
  \urlprefix\url{https://link.aps.org/doi/10.1103/RevModPhys.80.787}

\bibitem{Soda-Sudhir22}
\ifmmode~\check{S}\else \v{S}\fi{}oda B, Sudhir V and Kempf A 2022 {\em Phys.
  Rev. Lett.\/} {\bf 128}(16) 163603
  \urlprefix\url{https://link.aps.org/doi/10.1103/PhysRevLett.128.163603}

\bibitem{Glaetze-Hammerer10}
Glaetzle A, Hammerer K, Daley A, Blatt R and Zoller P 2010 {\em Optics
  Communications\/} {\bf 283} 758--765 ISSN 0030-4018 quo vadis Quantum Optics?
  \urlprefix\url{https://www.sciencedirect.com/science/article/pii/S0030401809010554}

\bibitem{Ferreri-Domina19}
Ferreri A, Domina M, Rizzuto L and Passante R 2019 {\em Symmetry\/} {\bf 11}
  ISSN 2073-8994 \urlprefix\url{https://www.mdpi.com/2073-8994/11/11/1384}

\bibitem{Reina-Domina21}
Reina M, Domina M, Ferreri A, Fiscelli G, Noto A, Passante R and Rizzuto L 2021
  {\em Phys. Rev. A\/} {\bf 103}(3) 033710
  \urlprefix\url{https://link.aps.org/doi/10.1103/PhysRevA.103.033710}

\bibitem{Calajo-Rizzuto17}
Calaj\`o G, Rizzuto L and Passante R 2017 {\em Phys. Rev. A\/} {\bf 96}(2)
  023802 \urlprefix\url{https://link.aps.org/doi/10.1103/PhysRevA.96.023802}

\bibitem{Passante-Persico02}
Passante R and Persico F 2003 {\em Physics Letters A\/} {\bf 312} 319--323 ISSN
  0375-9601
  \urlprefix\url{https://www.sciencedirect.com/science/article/pii/S0375960103006789}

\bibitem{Vasile-Passante08}
Vasile R and Passante R 2008 {\em Phys. Rev. A\/} {\bf 78}(3) 032108
  \urlprefix\url{https://link.aps.org/doi/10.1103/PhysRevA.78.032108}

\bibitem{Messina-Vasile10}
Messina R, Vasile R and Passante R 2010 {\em Phys. Rev. A\/} {\bf 82}(6) 062501
  \urlprefix\url{https://link.aps.org/doi/10.1103/PhysRevA.82.062501}

\bibitem{Haakh-Henkel14}
Haakh H~R, Henkel C, Spagnolo S, Rizzuto L and Passante R 2014 {\em Phys. Rev.
  A\/} {\bf 89}(2) 022509
  \urlprefix\url{https://link.aps.org/doi/10.1103/PhysRevA.89.022509}

\bibitem{Passante-Vinci96}
Passante R and Vinci N 1996 {\em Physics Letters A\/} {\bf 213} 119--124 ISSN
  0375-9601
  \urlprefix\url{https://www.sciencedirect.com/science/article/pii/0375960196001442}

\bibitem{Rizzuto-Passante04}
Rizzuto L, Passante R and Persico F 2004 {\em Phys. Rev. A\/} {\bf 70}(1)
  012107 \urlprefix\url{https://link.aps.org/doi/10.1103/PhysRevA.70.012107}

\bibitem{Armata-Vasile16}
Armata F, Vasile R, Barcellona P, Buhmann S~Y, Rizzuto L and Passante R 2016
  {\em Phys. Rev. A\/} {\bf 94}(4) 042511
  \urlprefix\url{https://link.aps.org/doi/10.1103/PhysRevA.94.042511}

\bibitem{Noto-Messina14}
Noto A, Messina R, Guizal B and Antezza M 2014 {\em Phys. Rev. A\/} {\bf 90}(2)
  022120 \urlprefix\url{https://link.aps.org/doi/10.1103/PhysRevA.90.022120}

\bibitem{Power-Zienau59}
Power E~A and Zienau S 1959 {\em Philosophical Transactions of the Royal
  Society of London A: Mathematical, Physical and Engineering Sciences\/} {\bf
  251} 427--454 ISSN 0080-4614
  \urlprefix\url{http://rsta.royalsocietypublishing.org/content/251/999/427}

\bibitem{Compagno-Passante83}
Compagno G, Passante R and Persico F 1983 {\em Physics Letters A\/} {\bf 98}
  253--255 ISSN 0375-9601
  \urlprefix\url{https://www.sciencedirect.com/science/article/pii/0375960183908642}

\bibitem{Armata-Kim17}
Armata F, Kim M~S, Butera S, Rizzuto L and Passante R 2017 {\em Phys. Rev. D\/}
  {\bf 96}(4) 045007
  \urlprefix\url{https://link.aps.org/doi/10.1103/PhysRevD.96.045007}

\end{thebibliography}
\end{document}